# JOINUS:
# A User-friendly Open-source Software to Simulate Digital Superconductor Circuits

Sasan Razmkhah and Pascal Febvre

*Abstract*—Single flux quantum (SFQ) based circuits are the subject of renewed attention due to their high speed and their very high energy efficiency. However, the need of cryogenic temperature, the complex physics of Josephson junctions and the lack of proper EDA tools causes slow progress in the field of superconducting electronics. In this work we introduce a new open source program named JOsephson INterface Utility Software (JOINUS [1]) that incorporates SPICE-based simulator engines, improved physical models and several powerful built-in routines. JOINUS is based on a user-friendly environment available on Linux, MacOS and Windows platforms to simplify the design and analysis of superconducting digital circuits.

*Index Terms*—Superconducting electronics, Single Flux Quantum, SFQ, Josephson Simulator, Graphical User Interface, EDA, TCAD

## I. INTRODUCTION

IN recent years the integration of semiconductor-based logic has slowed down as the technology is getting closer to limits associated to speed and overwhelming power consumption for very dense circuits. To overcome the present limitations, several technologies are being studied since there is a need not only for new much more energy-efficient logic circuits but also for different approaches in terms of methods of computations, departing from the widely spread von-Neumann architecture. In this context superconductor circuits are a strong candidate for advancing the capabilities of CMOS circuits for some applications [2]. Because of the non-dissipative nature of superconductors, SFQ-based circuits exhibit very high energy efficiency simultaneously with very high speed. However, the complex physics surrounding superconductors and their main active element, e.g. the Josephson junction, the operation at high frequencies in the microwave range, and the use of a dynamic pulsed logic, make it difficult to get familiar with the operation of SFQ circuits. Therefore, computer-based design tools are strongly needed, not only for enabling the development of the technology, but also for educational purposes.

The problem faced by the currently available design tools is that they are generally not all openly available, their interface is not always user-friendly and, more importantly, there is no general software package comprising all softwares necessary to design SFQ circuits from A to Z. At last the available tools do not have the accuracy and the functionalities needed to develop complex models or novel devices. Consequently, newcomers must spend much time to get used to the design, handle non compatibilities of formats for instance, and gather different tools to achieve each design task. As a result, each team has developed its own tools. This has scattered the efforts instead of focusing them on the development of a generic software suite for everyone. This problem has been tackled recently in some projects such as Cryogenic Complex Computing [3] and ColdFlux SuperTools [4].

## II. STRUCTURE OF SOFTWARE

In this context we have developed an interface named JOsephson INterface Utility Software (JOINUS) for SPICE-based simulation engines such as JSIM [5] or JoSIM [6]. This interface can later include more simulators such as WRSPICE, an improvement of SPICE3 [7], or PSCAN [8]. It operates with both voltage-based or phased-based simulators. The main goal of the program was to create an environment to simplify the process of creating netlists, running simulations, optimizing circuits, verifying the influence of some parameters, storing and displaying simulation results. A circuit optimizer and a margin and yield analysis tool are also under development for later releases of JOINUS. However, the program has also the possibility to take thermal effects into account in the circuit netlist by adding automatically Johnson noise to resistors. It features some other functions like the automatic simulation of *I-V* curves, and it can perform different kinds of parametric analysis. It can also simulate circuits at different temperatures, away from the nominal 4.2K temperature. The program was developed with the Qt integrated development environment [9] and can therefore operate on all platforms.

JOINUS is open-source and can be modified, duplicated and distributed for free. It allows future users to change simulation parameters, adapt the software for different materials and processes, or add new models and capabilities.

### A. Graphical user interface

JOINUS is object-oriented and written in the C++ language, its interface is shown in Figure 1 while Figure 2 displays the flowchart of the execution process. The parameters and netlist objects are loaded and displayed in the front panel upon software launch, while the netlist is analyzed and modified according to the options selected in the left panel. After selecting the parameters and pressing the start button, the program processes the input data and the netlist. If no problem is detected, the chosen simulation method runs by calling the desired engines (JSIM or JoSIM, with or without noise taken into account) and generates output. Otherwise the error is indicated in the console in the bottom right part of the interface.

S. Razmkhah and P. Febvre are with the IMEP-LAHC-CNRS, Université Savoie Mont Blanc, 73376 Le Bourget du Lac, France (e-mail: sasan.razmkhah@univ-smb.fr, pascal.febvre@univ-smb.fr).

The data are plotted if the user has selected the plotting option. Either an open-source plotting software or the built-in JOINUS plotter can be used. If the simulation is successful, the result can be saved in the directory corresponding to the selected path.

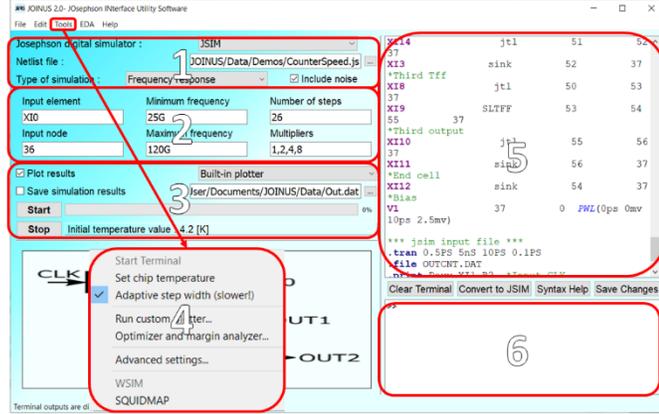

Figure 1. Interface of the JOINUS software. 1) Options and selection of simulation methods. 2) Parameter window for simulations. 3) Choice of output generation and display. 4) Links to stand-alone applications and settings. 5) Netlist editor and syntax check. 6) Console to display outputs and errors.

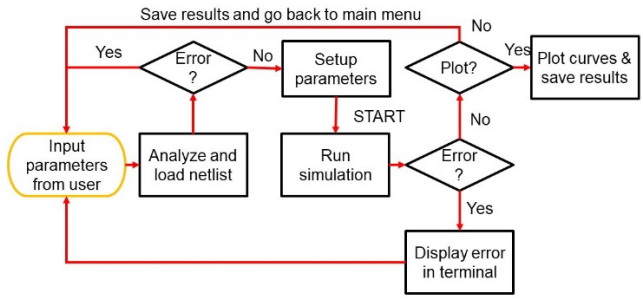

Figure 2. Flowchart of the JOINUS execution process.

## B. Plotting graphs

One of the user-friendly features of JOINUS is its ability to directly plot data, with different open-source programs such as GNUplot and XMGrace, or with a powerful built-in plotter which supports different types of data such as ASCII and CSV. The JOINUS built-in plotter has all the options included in conventional plotters and is directly linked to the main program. Therefore, it is aware of the type of required simulations and adapts its settings accordingly to display output results in the most appropriate way. An example is shown in Figure 3.

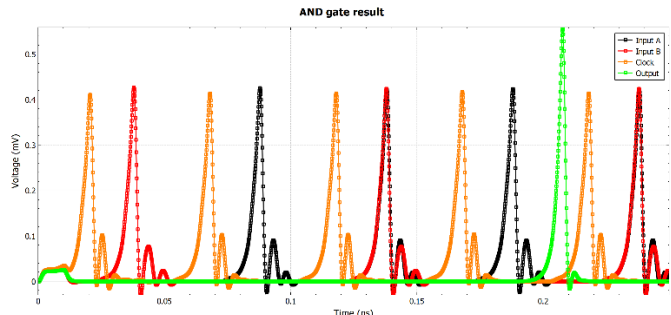

Figure 3. Input and output pulses shown in the JOINUS built-in plotter for an AND gate in SFQ technology.

## III. SOFTWARE FEATURES

Two models are available for the Josephson junction in JSIM [5] and JoSIM [6] simulators. The second model, which comprises a piece-wise linear conductance curve, requires the gap voltage value obtained from the generalized energy-gap equations at finite temperatures [10] from the BSC theory. The gap voltage depends also on the current running in the superconductor and is calculated from the equations of diffusion developed by Usadel [11].

### A. Influence of temperature

For most circuits we can consider the bulk limit boundary condition and assume that the superconductor material has a symmetric lattice such as for elements like tin and aluminum [10]. The elemental bulk superconductor's lattice structure is homogenous and isotropic. In this case, the bandgap energy can be estimated with (1) by fitting the temperature-dependent energy gap integral form (see equation 3.53 of [10]):

$$\Delta = \Delta_0 \sqrt{\cos\left(\frac{\pi}{2}(T/T_C)^2\right)} \quad (1)$$

where T is the temperature of the superconductor, $T_c$ its critical temperature and

$$2\Delta_0 = 3.53 k_B T_C. \quad (2)$$

Other approximations of $\Delta$, like for instance the one given in [11], have been derived. They give values similar to the ones obtained with equation (1) within about 1%. The value of the gap energy at 0K can be estimated with (2). $\Delta_0$ is the energy that an electron needs to go from condensate (superconductor) state to the normal state at 0K. Hence the gap energy of Cooper pairs is $2\Delta_0$. The gap voltage is calculated from the bandgap energy with (3):

$$v_g = \frac{2\Delta}{e} \quad (3)$$

The critical current of superconductor and Josephson junctions is also a function of temperature. It is automatically adjusted in JOINUS. The critical current variations with temperature are estimated by considering the Helmholtz's free energy and the changes in penetration depth due to Cooper pairs breaking [10], [12].

$$I_C = I_0(1 - (T/T_C)^2)\sqrt{1-(T/T_C)^4} \quad (4)$$

Here, $I_0$ is the critical current of the junction at 0K, in other words, it is the value of current that has enough energy to break the Cooper pairs and cause voltage across the Josephson junction's terminals at the fundamental level at the absolute temperature. The normal resistance of the junction is estimated from (5) [13].

$$R_N = \frac{\pi \Delta_0}{2eI_0} \times \tanh\left(\frac{\Delta_0}{2k_BT}\right) \quad (5)$$

These equations are pre-processed by JOINUS in the netlist of the circuit under study. Figure 4 shows a simulation result for a Josephson AND gate at 0, 3.5 and 7K obtained by including thermal noise. Since RSFQ circuits are based on a pulse logic, the Josephson AND gate generates a pulse signal at the output only if two pulses (one for input A and one for input B) arrive between two clock pulses. Figure 4 shows that the delay between the clock and the output pulses decreases as the temperature increases. The output pulse shape is also slightly modified. By increasing the temperature to a certain point we cause the circuit to start oscillating, causing a malfunction of the cell, as is seen at the temperature of 7K.

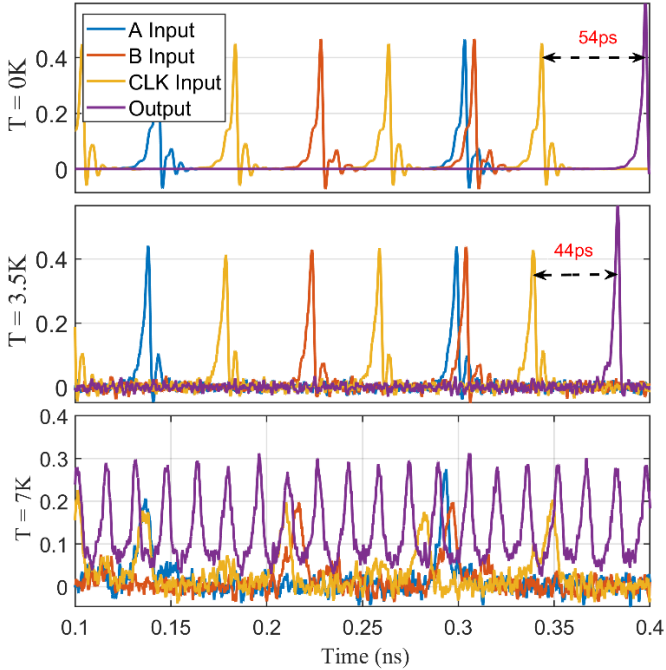

Figure 4. Simulation of a Josephson AND gate for three different temperatures. Simulations show the change in the timing of the output pulse with respect to the clock pulse, and the malfunction of the cell as temperature increases at 7K.

### B. Influence of thermal noise

JOINUS uses SPICE-based simulators, therefore Josephson junctions are described as lumped elements. The noise model used for resistors is based on Johnson noise [14], [15], [16] and added as a noise current source in parallel with the resistor. Its expression per unit frequency bandwidth depends on the temperature and value of the resistor $R$ as shown in (6):

$$\langle I_N \rangle = \sqrt{\frac{4k_B T}{R}}. \quad (6)$$

In practice JOINUS detects the resistors of the netlist and calculates the noise power based on the resistor and temperature then it adds the noise current source. To give the user the possibility to exclude the noise source from some of the resistors JOINUS does not add noise sources to resistors whose name includes the "nonoise" expression. For example, if the user wants to get an output over a load resistor without adding noise, it can simply name it "Rnonoise" or "Rnonoise-load".

The total noise current value used by JSIM depends on the step size of the transient analysis. This is the case in most simulators where the transient steps determine the bandwidth, hence the energy of the white noise. By getting enough samples and averaging them, we can obtain more accurate noise simulation results. We discuss this point in more details in the "current-voltage characteristics" section.

It is important to mention that simulations are dependent on the simulation step and can give results which are not physical. For instance with .TRAN=0.1 ps the corresponding noise bandwidth is 10 THz which is much above the pair breaking frequency. This can lead to simulated results which do not correspond to experimental ones. A more correct approach to the thermal noise is to use the random walk or Brownian distribution [17] since the thermal noise results physically from the random movement of the quasi-electrons, especially when the superconductor material is at the dirty limit.

### C. Current-voltage characteristics

Most of simulators are designed to do time-transient analysis. If other types of simulations are needed, they must be implemented manually through dedicated scripts written by the user, which is time consuming. JOINUS includes a few automated routines that not only ease the design of digital circuits, but also allow the user to create new devices and mixed-signal circuits. For instance $\pi$- and $\varphi$-junctions can be used to account for the properties of magnetic barriers or of s-wave/d-wave interfaces. Passive RF elements can also be incorporated through dedicated compact models.

One of the built-in functions is the simulation of $I$-$V$ curves from the circuit netlist. There are various ways to calculate the $I$-$V$ characteristics.

a) The simplest way is to apply a fixed current, simulate the netlist and calculate the average voltage. Then create a new netlist where the value of the DC current source is incremented and continue this for the desired number of data points. With this method the current is increased each time from the same initial (usually null) value. This method does not correspond to the real dynamics followed by the Josephson junction when the current is swept experimentally. In particular the hysteretic behavior observed experimentally in some cases cannot be reproduced since the initial conditions of simulations are different. Also the simulation time is quite long in this case since one full time-domain simulation is needed for each point in the $I$-$V$ curve.

b) A better method consists of simulating the circuit by doing the simulations in similar conditions as the ones met during measurements. This is done by providing a triangle current signal at the circuit input and calculating the averaged output voltage. Contrary to the first method there is no loss of the initial state of the junction hence the hysteretic behavior can be simulated. Also the simulation corresponds to a single netlist and is much faster. Nevertheless, since a junction oscillates at the Josephson frequency, proportional to the DC average voltage, when a bias current higher than the critical current $I_c$ is applied, it is not possible to obtain the DC voltage value directly for each input current. In practice the DC voltage is obtained by averaging the oscillating voltage over a sufficient duration of many periods for good accuracy. The easiest way to do that is to increase the applied current by steps and wait long enough at each step to obtain an accurate average of the oscillating voltage (see Figure 5).

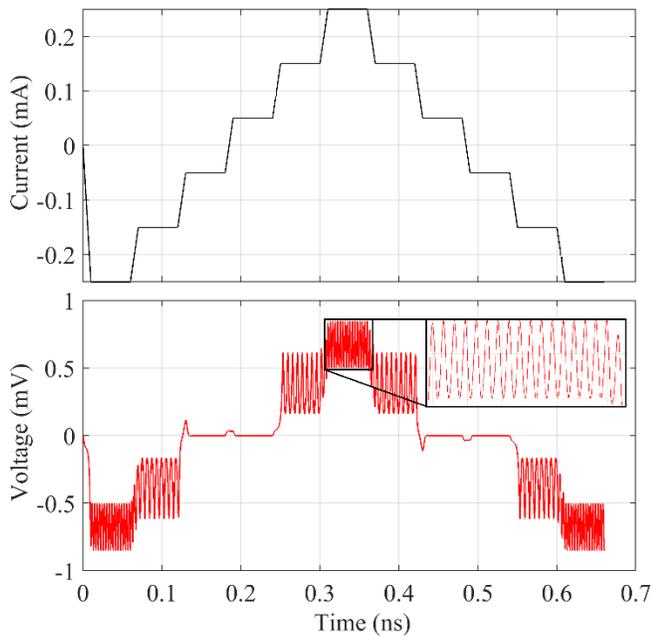

Figure 5. *I-V* curve calculation method used in JOINUS. The input current is increased by steps so that the voltage can be averaged over each step duration corresponding to a fixed current.

While this method is successfully used in experiments, it is not always adequate for simulations since the frequency of oscillations can vary tremendously while the applied bias current is swept: this is the case in particular just above the critical current where the DC voltage departs from the initial zero value. Indeed the longest periods of Josephson oscillations correspond to the lowest DC voltages. Consequently this region in the *I-V* curve requires long simulation times to make sure that each DC voltage value is accurate over averaging, even though long averaging times are not necessary for higher voltages that correspond to lower periods of oscillation. In practice, this problem does not happen experimentally since any measurement that is done takes a long time with respect to the period of oscillations in the picosecond range, averaging naturally the signal as a consequence, even in the region just above *Ic*.

c) From the previous analysis a better method to reduce further the duration of simulations consists of adapting the duration of the averaging steps to match the period of oscillations for each current bias point in the *I-V* curve. Therefore JOINUS features also an adaptive simulation method for which the averaging duration is chosen dynamically according to the shape of the *I-V* curve. In more critical points where the change of the *I-V* curve slope is stronger, e.g. when the absolute value of the second derivative of the *I-V* curve is high, the averaging time is chosen to be longer, as shown in Figure 6. To do so, JOINUS first runs a fast but inaccurate *I-V* calculation with the netlist. Then it processes the shape of the *I-V* curve and creates a vector of time stamps from which a new netlist is built upon. The new netlist is simulated with averaging durations for each bias current point based on the time stamps vector. The result of this method of calculation is shown in Figure 7. It gives an *I-V* characteristics that is closer to the analytical solution described in [17].

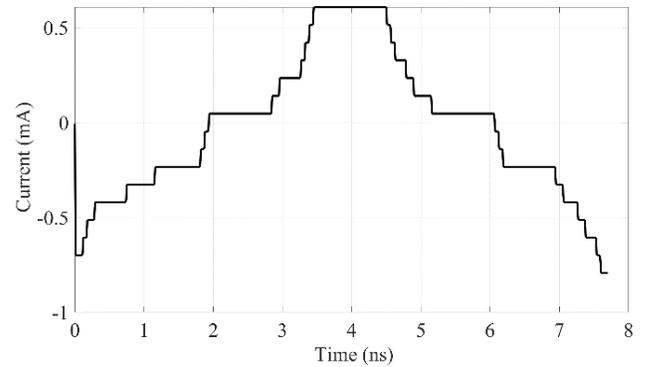

Figure 6. Example of improved bias current sweep of a Josephson junction. The averaging time is modified dynamically to draw the *I-V* characteristics. This method is more accurate in presence of thermal noise which modifies the frequency of voltage oscillations and consequently the optimum duration for voltage averaging.

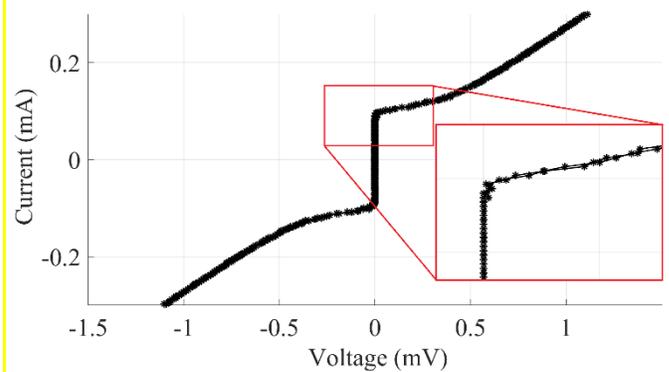

Figure 7. *I-V* characteristics of a damped junction with $I_C$ = 110µA simulated with an averaging time modified dynamically.

We give below a few examples that show that the algorithms implemented in JOINUS allow one to simulate with accuracy *I-V* curves exhibiting several physical manifestations expected with Josephson junctions.

*1) Influence of noise*
In JSIM, the noise is simulated as a random current source whose bandwidth is limited by the simulation transition timestep. As seen in Figure 8, the normal averaging method described above, with fixed time averaging steps, gives a curve that is different from measured data. This is due to the fact that, for each given bias current point, the effective current applied to the junction varies for each time step associated to the simulation of the output voltage during the averaging period. It is indeed the sum of the applied DC current and of the random thermal noise current. Consequently the frequency of simulation fluctuates as well as the resulting average voltage, which we clearly see on the graph of the top of Figure 8. A remedy consists of increasing the averaging time, which corresponds to more sample points and leads to a noise closer from the expected Gaussian distribution. But this comes at the price of a much longer simulation time.
By using the adaptive method described above, the optimum time for simulation is found automatically for each current bias point, which leads to an *I-V* curve much more in agreement with the ones measured experimentally, as can be seen on the bottom graph of Figure 8. For *I-V* characteristics with 200 points, the adaptive method takes about 2.7s while to obtain the same result with the normal method, the simulation time was 7.3s. This method can be used between any two nodes of any circuit, simple or complex, as long as a Josephson *I-V* response exists between these nodes.

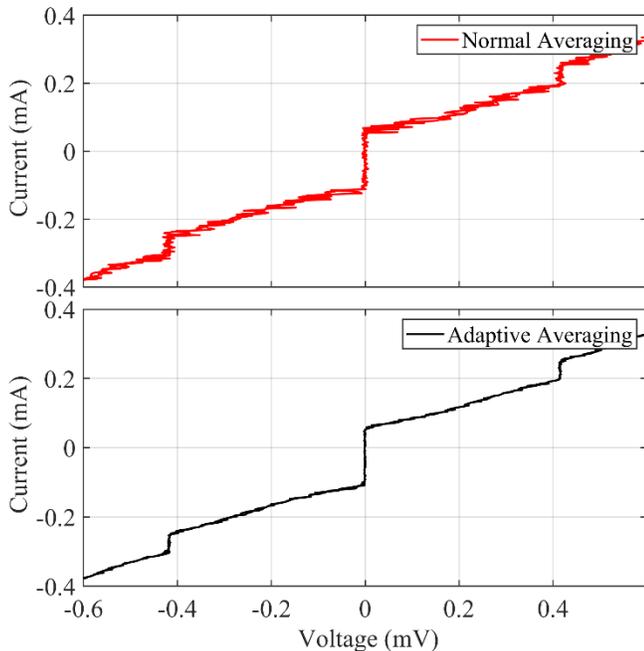

Figure 8. *I-V* characteristic of a damped Josephson junction with exaggerated noise and in presence of a microwave signal. The normal method (simulation time = 3.7s) uses fixed steps while the adaptive method (simulation time = 16.4s) uses longer time steps for averaging.

*2) Influence of external microwave signals*
When a Josephson junction is irradiated with a microwave signal, its *I-V* curve is modified since the applied signal induces additional tunneling currents depending on the DC bias point. The changes on the Cooper pair DC current for specific voltages is known as Shapiro steps [18] which are due to signal mixing by the non-linear dynamics of the junction when the applied microwave signal frequency is equal to the Josephson frequency of the internal oscillating signal. Shapiro steps appear for DC voltages whose corresponding Josephson frequency equals the microwave frequency or his harmonics. Since Shapiro steps are connected to the Cooper pair current, they can be simulated with Josephson simulator engines. Figure 9 shows the *I-V* characteristics of a single Josephson junction in presence of microwave radiations at different microwave frequencies.

*3) Influence of the Stewart-McCumber parameter*
The Stewart-McCumber parameter [19], also called damping factor, is a characteristics of a Josephson junction that is connected with the hysteresis of *I-V* characteristics. This parameter is very important to design logic circuits since it determines the power consumption and speed of circuits. Figure 10 shows the *I-V* curve of a junction with different damping factors. The hysteresis effect expected for values of the damping factor higher than about 2 are observed in simulations.

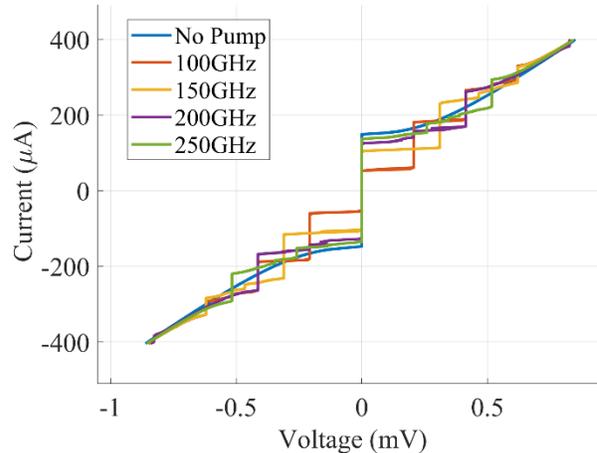

Figure 9. *I-V* characteristics of a Josephson junction simulated by JOINUS in presence of microwave signals of different frequencies. The heights of the Shapiro steps depend on the power of the microwave signal, while their voltage is directly connected to the signal frequency through the Josephson relation.

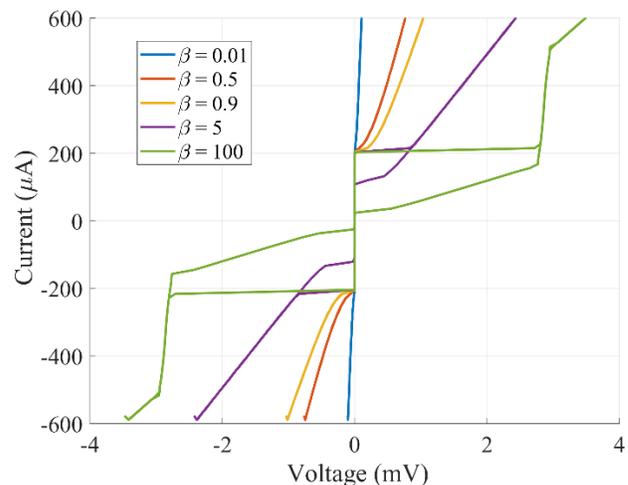

Figure 10. Simulated *I-V* curves of a Josephson junction with different damping factors.

*D. Parametric analysis*
JOINUS also include routines for parametric and temperature analysis. The parametric analysis is different from the temperature analysis since temperature is not a parameter that is explicitly included in netlists. Also temperature affects several parameters of the netlist simultaneously. To perform the analysis JOINUS automatically generates different netlists based on the parameter that the user wants to sweep, simulates all netlists and generates outputs for each of the steps. The user can directly monitor the changes of output results on the JOINUS built-in plotting window by sliding a cursor connected to the parameter under study. The parameter analysis functionality is demonstrated in Figure 9 and Figure 10 as the input microwave signal frequency and the shunt resistor of the Josephson junction are swept, respectively.

## E. Tests of the frequency response of digital circuits

JOINUS can find automatically the maximum frequency of operation of a digital circuit. It generates a pulse train at the input of the circuit based on the oscillation of an overbiased Josephson junction at frequencies increasing with time, which is done by changing the bias current with time. JOINUS simulates the output signal by comparing the number of output pulses with the expected number of pulses. The maximum frequency of the circuit is determined when the two figures deviate from each other. However, this function is currently only working with circuits whose output(s) can be determined by a single input. In future versions a truth table that supports any type of circuits will be implemented. An example is shown in Figure 11 for a 3-bit counter.

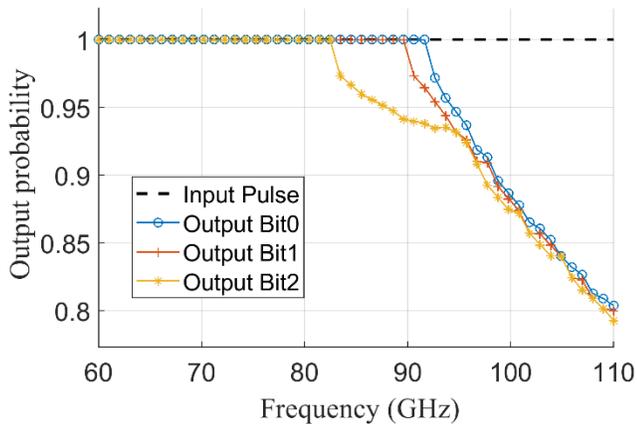

Figure 11 Frequency response of a 3-bit pulse counter circuit simulated without noise. The pulse number is normalized and digitization error is eliminated from results.

By combining the parameter analysis and the maximum frequency calculation, the change in the frequency response of a TFF cell can be calculated as the bias value of the cell is shifted. The result is shown in Figure 12.

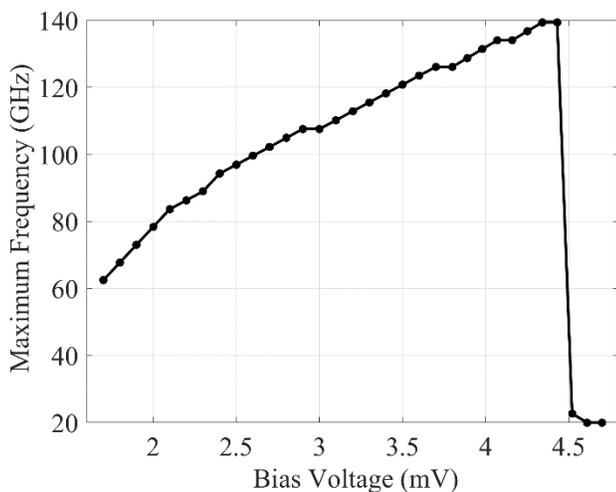

Figure 12 Maximum response frequency of a TFF cell versus input voltage bias of the cell.

## F. Bit Error Rate Analysis

JOINUS can also simulate automatically the Bit Error Rate (BER) of circuits. The BER function takes a digital input sequence from the user for a given parameter to study, simulates the output many times and compares it with the expected output associated to a circuit operating correctly. The percentage of failures is calculated for this parameter, which is swept over a range defined beforehand by the user. JOINUS then plots the BER graph at the end of the parameter analysis, as is shown in Figure 13.

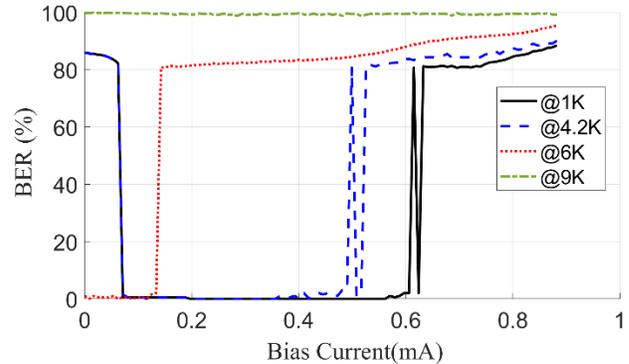

Figure 13. Bit Error Rate calculation at different temperatures for a JTL cell.

## IV. CONCLUSION

We have developed a program named JOINUS to ease the process of editing and simulating SFQ-based circuits. This program brings together different Josephson simulator engines such as JSIM and JoSIM, simulates netlists through different built-in routines, saves data in user-defined format and plots results. JOINUS has also the ability to modify netlists to add temperature and noise effects. Different functions are built within JOINUS such as current-voltage characterization and parametric analysis. Additions of further capabilities are in progress.

## ACKNOWLEDGMENT

The research is based upon work supported by the Office of the Director of National Intelligence (ODNI), Intelligence Advanced Research Projects Activity (IARPA), via the U.S. Army Research Office grant W911NF-17-1-0120. The views and conclusions contained herein are those of the authors and should not be interpreted as necessarily representing the official policies or endorsements, either expressed or implied, of the ODNI, IARPA, or the U.S. Government. The U.S. Government is authorized to reproduce and distribute reprints for Governmental purposes notwithstanding any copyright notation herein. We thank Coenrad Fourie and Johannes Delport from Stellenbosch University for stimulating discussions about JoSIM software.

## REFERENCES

[1] "Superconductor-Electronics - Overview," *GitHub*. [Online]. Available: https://github.com/Superconductor-Electronics. [Accessed: 19-May-2019].
[2] "International Roadmap for Devices and Systems (IRDS[TM]) 2018 Edition - IEEE International Roadmap for Devices and Systems[TM]." [Online]. Available: https://irds.ieee.org/editions/2018. [Accessed: 17-Jul-2019].
[3] M. A. Manheimer, "Cryogenic Computing Complexity Program: Phase 1 Introduction," *IEEE Trans. Appl. Supercond.*, vol. 25, no. 3, pp. 1–4, Jun. 2015, doi: 10.1109/TASC.2015.2399866.
[4] C. J. Fourie *et al.*, "ColdFlux Superconducting EDA and TCAD Tools Project: Overview and Progress," *IEEE Trans. Appl. Supercond.*, vol. 29, no. 5,

pp. 1–7, Aug. 2019, doi: 10.1109/TASC.2019.2892115.
[5] E. S. Fang and T. Van Duzer, "A Josephson integrated circuit simulator (JSIM) for superconductive electronics application.," *Ext Abstr 1989 Int Supercond Electron Conf ISEC 89*, pp. 407–410, 1989.
[6] J. A. Delport, K. Jackman, P. l Roux, and C. J. Fourie, "JoSIM—Superconductor SPICE Simulator," *IEEE Trans. Appl. Supercond.*, vol. 29, no. 5, pp. 1–5, Aug. 2019, doi: 10.1109/TASC.2019.2897312.
[7] S. R. Whiteley, "Josephson junctions in SPICE3," *IEEE Trans. Magn.*, vol. 27, no. 2, pp. 2902–2905, Mar. 1991, doi: 10.1109/20.133816.
[8] S. Polonsky, P. Shevchenko, A. Kirichenko, D. Zinoviev, and A. Rylyakov, "PSCAN'96: new software for simulation and optimization of complex RSFQ circuits," *IEEE Trans. Appl. Supercond.*, vol. 7, no. 2, pp. 2685–2689, Jun. 1997, doi: 10.1109/77.621792.
[9] T. Q. Company, "Qt | Cross-platform software development for embedded & desktop." [Online]. Available: https://www.qt.io. [Accessed: 18-Feb-2019].
[10] M. Tinkham, *Introduction to Superconductivity*. Courier Corporation, 1996.
[11] S. A. Sergeenkov, "Electric field dependence of the thermal conductivity of a granular superconductor: Giant field-induced effects predicted," *J. Exp. Theor. Phys. Lett.*, vol. 76, no. 3, pp. 170–174, Aug. 2002, doi: 10.1134/1.1514762.
[12] T. Van Duzer and C. W. Turner, "Principles of superconductive devices and circuits," 1981.
[13] W. Haberkorn, H. Knauer, and J. Richter, "A theoretical study of the current-phase relation in Josephson contacts," *Phys. Status Solidi A*, vol. 47, no. 2, pp. K161–K164, Jun. 1978, doi: 10.1002/pssa.2210470266.
[14] J. B. Johnson, "Thermal Agitation of Electricity in Conductors," *Phys. Rev.*, vol. 32, no. 1, pp. 97–109, Jul. 1928, doi: 10.1103/PhysRev.32.97.
[15] H. Nyquist, "Thermal Agitation of Electric Charge in Conductors," *Phys. Rev.*, vol. 32, no. 1, pp. 110–113, Jul. 1928, doi: 10.1103/PhysRev.32.110.
[16] D. T. Gillespie, "The mathematics of Brownian motion and Johnson noise," *Am. J. Phys.*, vol. 64, no. 3, pp. 225–240, Mar. 1996, doi: 10.1119/1.18210.
[17] V. Ambegaokar and B. I. Halperin, "Voltage Due to Thermal Noise in the dc Josephson Effect," *Phys. Rev. Lett.*, vol. 22, no. 25, pp. 1364–1366, Jun. 1969, doi: 10.1103/PhysRevLett.22.1364.
[18] S. Shapiro, "Josephson Currents in Superconducting Tunneling: The Effect of Microwaves and Other Observations," *Phys. Rev. Lett.*, vol. 11, no. 2, pp. 80–82, Jul. 1963, doi: 10.1103/PhysRevLett.11.80.
[19] W. C. Stewart, "Current-voltage characteristics of josephson junctions," *Appl. Phys. Lett.*, vol. 12, no. 8, pp. 277–280, Apr. 1968, doi: 10.1063/1.1651991.